\documentclass[sigconf]{aamas} 

\usepackage{balance} % for balancing columns on the final page
\usepackage{bm}
\usepackage{xcolor}
\setcopyright{none}
\acmConference[ALA '26]%
{Proc.\@ of the Adaptive and Learning Agents Workshop (ALA 2026)}%
{May 25 -- 26, 2026}%
{Paphos, Cyprus, https://alaworkshop2026.github.io/}%
{Aydeniz, Delgrange, Mohammedalamen, Yang (eds.)}%
\copyrightyear{2026}
\acmYear{2026}
\acmDOI{}
\acmPrice{}
\acmISBN{}
\acmSubmissionID{65}

\title[Convention Play]{\textit{ConventionPlay}: Capability-Limited Training for Robust \\ Ad-Hoc Collaboration}

\author{Abhishek Sriraman}
\affiliation{
  \institution{University of Sheffield}
  \city{Sheffield}
  \country{United Kingdom}}
\email{asriraman1@sheffield.ac.uk}

\author{Eleni Vasilaki}
\affiliation{
  \institution{University of Sheffield}
  \city{Sheffield}
  \country{United Kingdom}}
\email{e.vasilaki@sheffield.ac.uk}

\author{Robert Loftin}
\affiliation{
  \institution{University of Sheffield}
  \city{Sheffield}
  \country{United Kingdom}}
\email{r.loftin@sheffield.ac.uk}

\begin{abstract}
Ad-hoc collaboration often relies on identifying and adhering to shared conventions. However, when partners can follow multiple conventions, agents must do more than simply adapt; they must actively steer the team toward the most effective joint strategy. We present \textit{ConventionPlay}, a reinforcement learning-based approach that extends cognitive hierarchies to include a diverse population of adaptive followers. By training against partners with varied capability limits, our agent learns to probe its partner's repertoire, leading the team when possible and following when necessary. Our results in canonical coordination tasks show that \textit{ConventionPlay} achieves superior coordination efficiency, particularly in settings where conventions have differentiated payoffs.
\end{abstract}

\keywords{Ad-Hoc Collaboration; Multi-Agent Reinforcement Learning}

\newcommand{\BibTeX}{\rm B\kern-.05em{\sc i\kern-.025em b}\kern-.08em\TeX}

\begin{document}
\pagestyle{fancy}
\fancyhead{}

\maketitle 

%%%%%%%%%%%%%%%%%%%%%%%%%%%%%%%%%%%%%%%%%%%%%%%%%%%%%%%%%%%%%%%%%%%%%%%%
\section{Introduction} 

Cooperative multi-agent tasks often admit a variety of different solutions, with success depending on the agents' ability to coordinate their strategies.  In such tasks, joint strategies can exhibit different levels of compatibility, with performance degrading when individual strategies are misaligned.  We describe collections of mutually compatible strategies as ``conventions''.  Within the broader challenge of designing agents to collaborate with unseen teammates \cite{mirsky2022surveyadhocteamwork}, the ability to recognize and align with established conventions is crucial for effective ad-hoc collaboration \cite{Stone_Kaminka_Kraus_Rosenschein_2010,DBLP:conf/iclr/ShihSKES21,10.1145/3306618.3314268}.  In this work, we propose a reinforcement learning approach to training collaborative agents that go beyond simply adapting to their partner's convention.  Instead, when teamed with adaptive partners, our agents actively steer the interaction towards the most effective joint strategy in the repertoire of conventions that those partners are capable of following.

As a motivating example, consider the setting where two humans that have not previously interacted need to communicate with each other to solve a task.  They each have varying levels of proficiency with overlapping sets of languages.  At the start of their interaction, they are motivated to identify a single preferred language that allows them to successfully coordinate on the underlying task.  Here the languages can be viewed as "conventions", and aligning on the preferred language is part of the coordination task.  A human in this scenario may try a few different languages to identify the relative proficiencies of their partner.  The feedback they receive from this interaction -- correctness, or willingness of their partner to speak a certain language -- will inform the choice of a common language to use for the rest of the interaction.

% TODO: Check that the "team" terminology is consistent with the rest of the paper
A common approach taken in previous work on ad-hoc collaboration \cite{strouseFCPCollaboratingHumans2021, zhaoMEPMaximumEntropy2022, carrollUtilityLearningHumans2020} is to train collaborative agents through a two-step process that first attempts to train a diverse ``population'' of agents for the target task, and then trains a ``best-response'' agent capable of coordinating with any member of this population.  In the first phase, a population of \textit{teams} of agents that can solve the task together is generated---for example, all members of a team may speak a common language.  Members of a team are trained in self-play, and may be incompatible with members of other teams.  Agents in this initial population effectively act as ``leaders'' which choose a fixed convention and expect other agents to adapt to this convention.  In the second phase, a generalist agent is trained against this entire population, such that it learns to adapt to conventions followed by its partners---for example, a multi-lingual agent that adapts to whichever language its partner speaks.  Since these generalist agents are trained against partners who follow fixed conventions, they effectively become ``followers'' that adapt to their partner's convention.  A shortcoming of this approach is that such followers will fail to account for partners that are themselves capable of some degree of adaptation, and may not be limited to a single (potentially suboptimal) convention.

% TODO: Need to expand on the summary of the paper
In this work, we introduce \textit{ConventionPlay}, a reinforcement learning-based approach to training adaptive agents that are capable of coordinating with partners that may themselves be adaptive.  The novelty of this algorithm lies in a structured training population that includes a diverse set of \textit{generalist followers}---agents designed to adapt to specific, restricted subsets of conventions.  Exposing our final agent to these adaptive partners forces it to move beyond purely reactive behavior; it must actively probe a partner's behavior to decide when to lead the team toward an optimal strategy, and when to follow a partner's established convention.  Our experimental results demonstrate that, by explicitly training agents that are capable of both leading and following, \textit{ConventionPlay}  % If you use italics more than once for the algorithm name, you should use it everywhere
outperforms state-of-the-art methods in complex ad-hoc collaboration scenarios that involve multiple, mutually incompatible conventions.

%%%%%%%%%%%%%%%%%%%%%%%%%%%%%%%%%%%%%%%%%%%%%%%%%%%%%%%%%%%%%%%%%%%%%%%%
\section{Related Work}
To contextualize the development of \textit{ConventionPlay}, we review relevant literature in population diversity for ad-hoc collaboration, convention-free strategies, and partner shaping.

\paragraph{Population Diversity} The most common method for achieving robust ad-hoc collaboration is by anticipating the range of possible solutions to the task. A number of methods focus on computing a diverse set of solutions and training a generalist agent against this population. TrajeDI \cite{lupuTrajectoryDiversityZeroShot2021} achieves this by incentivizing policies to generate distinct trajectories. MEP \cite{zhaoMEPMaximumEntropy2022} encourages the population to cover the entire state-action space. LIPO \cite{charakorn2023generating} aims to learn policies that are incompatible with each other, yet individually optimal in self-play. Fictitious Co-Play (FCP) \cite{strouseFCPCollaboratingHumans2021} provides behavioral diversity by including checkpoints from the self-play training trajectory, simulating partners of varying skill levels. Our work focuses not on the discovery of these base conventions, but on leveraging the diversity of the base population to improve performance of a generalist agent. To that end, our proposed method is compatible with any of these diversity-generation techniques.

\paragraph{Convention-Free Strategies} Unlike methods that seek to generate diverse conventions, Other-Play \cite{huOtherPlay2020} attempts to eliminate the need for them entirely. It constructs "convention-free" policies by systematically avoiding arbitrary coordination choices, relying solely on the strategic structure of the game. While effective in abstract settings, we argue that avoiding conventions is often impossible in practice, particularly in Human-AI scenarios where social norms dictate behavior. Instead of avoiding conventions, our approach embraces them, aiming to construct agents capable of recognizing and coordinating with any convention present in a population. 

\paragraph{Partner Shaping} We first distinguish our work from the social influence approach of Jaques et al. \cite{pmlr-v97-jaques19a}. While they employ intrinsic rewards to explicitly encourage influence, our steering behavior emerges naturally as a solution to the meta-RL task of partner identification, avoiding the conflation of task success with auxiliary rewards. Similarly, LOLA \cite{foersterLOLALearningOpponentLearning2018} and M-FOS \cite{luModelFreeOpponentShaping2022} explore actions that influence partner behavior, but they are fundamentally designed to shape a partner's \textit{learning process} over time; they assume the partner's policy weights will shift in response to the agent's actions. In contrast, \textit{ConventionPlay} treats the partner's underlying policy as fixed but potentially multi-modal. Crucially, we do not attempt to model or access the partner's internal weights or learning parameters. Our steering actions are intended to \textit{probe} a partner's existing repertoire to identify and converge upon the most efficient shared convention.

%%%%%%%%%%%%%%%%%%%%%%%%%%%%%%%%%%%%%%%%%%%%%%%%%%%%%%%%%%%%%%%%%%%%%%%%
\section{Background}
\label{sec:background}

This section introduces the core formalisms required for our approach. We define the Dec-POMDP framework, formalize the concept of conventions, and describe the best-response and cognitive hierarchy models that form the basis of our training methodology.

\subsection{Dec-POMDPs}
% \todo{TODO: Generalize formalism to n-agent case, but clarify that we are working in n=2}. 

We model multi-agent tasks as a Decentralized Partially Observable Markov Decision Process (Dec-POMDP) \cite{oliehoekConciseIntroductionDecentralized2016}, defined by the tuple $\mathcal{M} = \langle S, A, P, R, \Omega, O, \gamma \rangle$. $S$ is the set of global states, $A = A_1 \times A_2$ is the joint action space, and $P(s_{t+1}| s_{t}, a_{t})$ defines the transition probabilities to a new state $s_{t+1}$ given the current state $s_t$ and joint action $a_t \in A$. The task is fully cooperative, with both agents receiving the same scalar reward $R(s_t, a_t)$. The joint observation $o_t \in \Omega = \Omega_1 \times \Omega_2$ is drawn according to the function $O(o_t|s_t, a_{t-1})$. 

An individual policy $\pi_i: \mathcal{H}_{i,t} \to \Delta(A_i)$ maps the agent's local action-observation history, $h_{i,t} = (o_{i,0}, a_{i,0}, o_{i,1}, \dots, a_{i,t-1}, o_{i,t}) \in \mathcal{H}_{i,t}$, to a probability distribution over its action space. A \emph{joint policy} $\pi = (\pi_1, \pi_2)$ specifies the behavior of both agents. The goal is to maximize the expected discounted return:
\begin{equation}
J(\pi) = \mathbb{E}_{s_0, a_t \sim \pi, s_{t+1} \sim P} \left[ \sum_{t=0}^\infty \gamma^t R(s_t, a_t) \right]
\label{eq:objective}
\end{equation}
where $\gamma \in [0, 1)$ is a discount factor. We denote the \emph{self-play} (SP) return as $J_{SP}(\pi_1) = J(\pi_1, \pi_1)$.  We assume throughout this work that each agent in $\mathcal{M}$ has identical action and observation spaces, such that an individual policy $\pi_i$ can control any agent $j$.

\subsection{Conventions}
We formalize a \textit{system of conventions} as a collection $\mathcal{C} = \{C_1, \dots, C_k\}$, where each convention $C_i \subset \Pi$ is a subset of the policy space. These sets are defined using two coordination margins, $\epsilon$ and $\delta$, which bound the performance of intra- and inter-convention pairings:

\begin{enumerate}
    \item \textbf{Intra-convention $\epsilon$-compatibility:} For any convention $C_i \in \mathcal{C}$, any two policies $\pi, \pi' \in C_i$ must satisfy:
    \begin{equation}
        J(\pi, \pi') \ge \min(J(\pi, \pi), J(\pi', \pi')) - \epsilon
    \end{equation}
    where $\epsilon \ge 0$ is a small bound. This implies that policies within the same convention are functionally compatible.

    \item \textbf{Inter-convention $\delta$-incompatibility:} For any two distinct conventions $C_i, C_j \in \mathcal{C}$ ($i \neq j$), any pair of individual policies $\pi \in C_i$ and $\pi' \in C_j$ must satisfy:
    \begin{equation}
        J(\pi, \pi') \le \min(J(\pi, \pi), J(\pi', \pi')) - \delta
    \end{equation}
    where $\delta > \epsilon$. The parameter $\delta$ represents a coordination gap, where a mismatch in conventions leads to a drop in utility.
\end{enumerate}
Together, these properties define conventions as compatibility classes in policy space. Importantly, two conventions $C_i$ and $C_j$ may yield identical self-play returns—$J_{SP}(\pi) = J_{SP}(\pi')$ for $\pi \in C_i, \pi' \in C_j$ — yet remain strictly $\delta$-incompatible.

\subsection{Best Response Training}
Given a finite \textit{set of training partners} $\mathcal{D}$, a best response $\pi$ is defined as the policy that maximizes the expected joint return:
\begin{equation}
\pi \in \text{BR}(\mathcal{D}) = \arg\max_{\pi \in \Pi} \mathbb{E}_{\pi' \sim \mathcal{D}} [J(\pi, \pi')]
\end{equation}
where the expectation is taken over a uniform distribution over $\mathcal{D}$. This objective ensures the agent optimizes for robust performance across the entire population. In our framework, these best responses are Bayes-adaptive; the agent treats the partner's identity $\pi' \in \mathcal{D}$ as a latent variable that must be inferred from the interaction history. Importantly, a generalist best response may achieve lower utility than a policy specialized for a single partner, reflecting the inherent cost of disambiguating the partner's strategy during the interaction.

\subsection{Cognitive Hierarchies}
The Cognitive Hierarchy (CH) \cite{Camerer2003} and K-level reasoning \cite{10.1257/aer.96.5.1737} models offer a scaffolding for structuring the policy space in multi-agent systems \cite{cuiSyKLRBRKlevelReasoning2021}. In this hierarchical approach, each level represents a leap in reasoning depth, where a level $k$ policy accounts for the behaviors of its lower-level counterparts. We utilize this hierarchy to establish a distribution of diverse agents, where each level defines a specific set of collaborative capabilities and assumptions. We denote individual level-$k$ policies as $\pi_k \in \mathcal{P}_k$ and sets of such policies as $\mathcal{D}_k$.

\textbf{Level-0 ($K_0$): Static Convention Policies.} The $K_0$ population, $\mathcal{P}_0$, comprises policies that each adhere to a fixed convention $C_i \in \mathcal{C}$. We denote a $K_0$ policy adhering to convention $i$ as $\pi_{0,i} \in C_i$. By definition, these agents are brittle: they expect a partner to match their specific mode of coordination and lack the flexibility to adapt to others. This level aligns with the populations generated by methods like MEP or TrajeDI; however, unlike SyKLRBR—which uses random $K_0$ policies to avoid conventions—we explicitly use $K_0$ to represent the diverse conventions present in ad-hoc scenarios.

\textbf{Level-1 ($K_1$): Adaptive Best Responses.} The $K_1$ population, $\mathcal{P}_1$, comprises policies defined as best responses to a specific distribution of $K_0$ partners: $\pi_1 \in \text{BR}(\mathcal{D}_1)$, where $\mathcal{D}_1 \subseteq \mathcal{P}_0$. While existing ad-hoc collaboration methods \cite{strouseFCPCollaboratingHumans2021, charakorn2023generating} typically produce generalist $K_1$ agents where $\mathcal{D}_1$ spans the entire $K_0$ population, our methodology allows for more granularity. Specifically, we first define a \textit{capability set} $\mathcal{K}_{\pi_1} \subseteq \mathcal{C}$ representing a restricted subset of conventions, and then define $\mathcal{D}_1 = \bigcup_{C \in \mathcal{K}_{\pi_1}} C$ as the corresponding set of $K_0$ policies. This allows us to model partners with varying degrees of flexibility. To quantify how effectively an agent adapts to its partner's specific limitations, we introduce the \textit{Coordination Efficiency} $\eta$ for a pair of agents $(\pi, \pi_k)$ as:
\begin{equation}
    \eta(\pi, \pi_k) = \frac{J(\pi, \pi_k)}{J^*(\pi_k)}
\end{equation}

where $J^*(\pi_k) = \max_{\hat{\pi} \in \Pi} J(\hat{\pi}, \pi_k)$ represents the maximum possible return achievable with partner $\pi_k$. For a static $K_0$ partner, $J^*(\pi_0) \approx J_{SP}(\pi_0)$. This metric allows us to assess whether an agent achieves the full potential of its partner's repertoire.

\textbf{Level-2 ($K_2$): Robust Ad-hoc Coordinators.} The $K_2$ agent is trained as a best response to the full diversity of the hierarchy: $\pi_2 \in \text{BR}(\mathcal{D}_2)$, where $\mathcal{D}_2 = \mathcal{P}_0 \cup \mathcal{P}_1$. By optimizing against this mixed population, the $K_2$ agent is incentivized to adopt a dual strategy: adhering to the fixed conventions of rigid $K_0$ partners while proactively signaling and steering toward optimal conventions when paired with flexible $K_1$ partners. To evaluate this capability, we define the optimal achievable return $J^*(\pi_1)$ for an adaptive partner as the maximum joint return possible with any policy belonging to a convention within that partner's repertoire:

\begin{equation}
    J^*(\pi_1) = \max_{C_i \in \mathcal{K}_{\pi_1}} \left[ \max_{\pi' \in C_i} J(\pi', \pi_1) \right]
\end{equation}

This formulation shifts the evaluation from global optimization to partner-aware coordination, testing whether an agent can realize the full potential of a partner's specific repertoire.

%%%%%%%%%%%%%%%%%%%%%%%%%%%%%%%%%%%%%%%%%%%%%%%%%%%%%%%%%%%%%%%%%%%%%%%%

\section{Method}

\begin{figure}[t]
    \centering
    \includegraphics[width=\columnwidth]{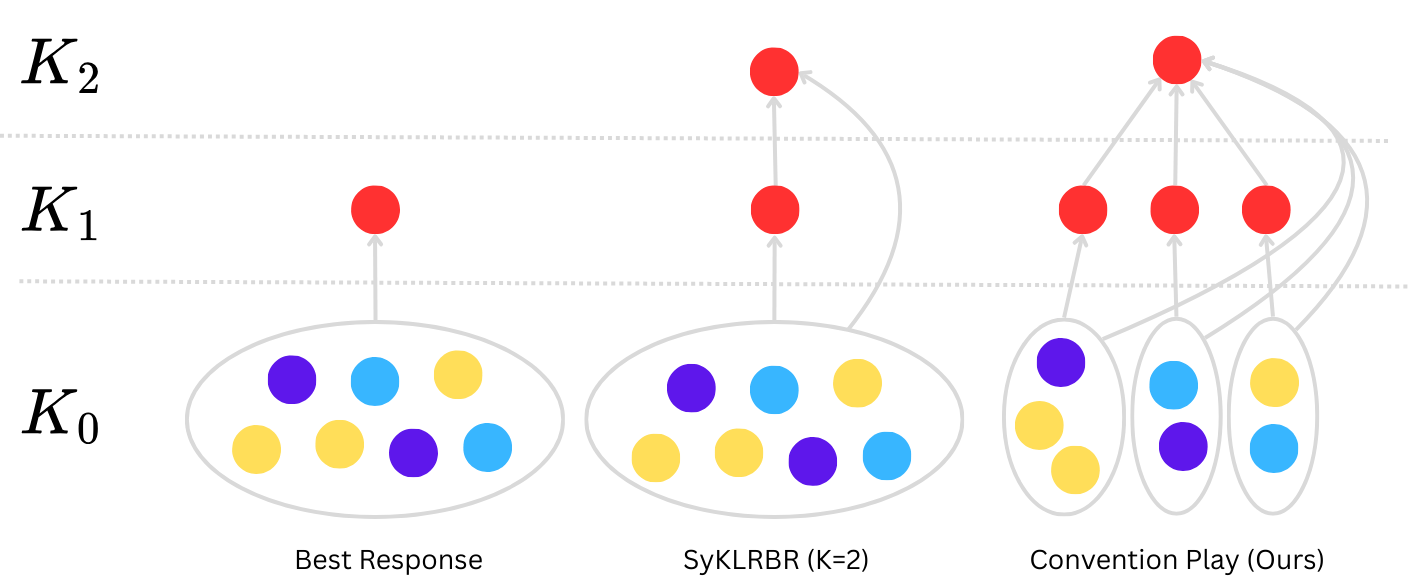}
    \caption{Comparison of the \textit{ConventionPlay} pipeline with existing ad-hoc collaboration methods. While baseline approaches diversify the $K_0$ population, we introduce diversity among $K_1$ followers to force the $K_2$ agent to infer its partner's limited repertoire and coordinate on the most effective shared convention.}
    \label{fig:convplay_diagram}
\end{figure}

This section provides an overview of the \textit{ConventionPlay} algorithm, and details on how the prerequisite $K_0$ and $K_1$ populations are created. \textit{ConventionPlay} trains an ad-hoc agent to be a best response to a diverse population of $K_0$ and $K_1$ partners. The algorithm follows three main steps: (1) generating a diverse set of $K_0$ policies to represent different base conventions; (2) training several $K_1$ agents, each adapted to specific subsets of the $K_0$ population; and (3) training the final $K_2$ agent to coordinate effectively across this entire hierarchy. The training hierarchy is illustrated in Figure~\ref{fig:convplay_diagram}.

At its core, \textit{ConventionPlay} relies on a $K_1$ population built from varied and restricted subsets of conventions. This structure forces the $K_2$ agent to transition from passive adaptation to active team steering. To coordinate effectively, the agent must employ probing strategies to reveal its partner's specific repertoire, allowing it to navigate the trade-offs between different coordination modes; by exposing the agent to partners that do not all support a single global optimum, we encourage the development of such a behavior. We describe the implementation of this hierarchy in the following sections, beginning with the generation of the base conventions.

\subsection{Generating a Base Population}

We generate our $K_0$ population by training MAPPO \cite{yuMAPPOSurprisingEffectiveness} across multiple random seeds. While specialized methods like TrajeDI or LIPO could also be used, we found that random initialization provided sufficient behavioral diversity. Notably, this approach does not guarantee that the population will partition into the well-structured system of incompatible policies described in our formalism; in practice, we may encounter complex, non-transitive compatibility structures driven in part by the task or learning architecture. We therefore verify this inter-convention incompatibility empirically to ensure our agents provide a clear foundation of distinct strategies for subsequent training levels.

\subsection{Capability-Aware Stratified Sampling}
\label{subsec:stratified_sampling}
To construct a $K_1$ population with diverse and limited capabilities, we define $M$ training subsets $\{\mathcal{D}_{1,1}, \dots, \mathcal{D}_{1,M}\}$ through stratified sampling of the $K_0$ population. The number of subsets $M$ is a hyperparameter that should ideally correspond to the number of conventions $|\mathcal{C}|$ in $K_0$. This method is designed to impose a ``capability ceiling'' on each $K_1$ agent---a property that is particularly critical in tasks with differentiated reward structures where conventions vary in objective value.  To generate these subsets, we first define the capability $\rho(\pi_{0,i})$ of an agent by its self-play return $J_{SP}(\pi_{0,i})$. We then identify $M$ anchor agents $\{\pi^*_{0,j}\}_{j=1}^M$ by selecting policies whose capabilities are closest to $M$ target performance levels linearly spaced across the range $[\min \rho, \max \rho]$. For each anchor, the associated training distribution $\mathcal{D}_{1,j}$ is formed by sampling from an eligibility pool $\mathcal{P}_{\text{elig}} = \{ \pi_{0,i} \in \mathcal{P}_0 : \rho(\pi_{0,i}) \le \rho(\pi^*_{0,j}) \}$. By including only policies that perform no better than the anchor, we ensure the anchor represents the performance ceiling within that $K_1$ agent's repertoire, forcing it to coordinate while respecting that limit.

%%%%%%%%%%%%%%%%%%%%%%%%%%%%%%%%%%%%%%%%%%%%%%%%%%%%%%%%%%%%%%%%%%%%%%%%
\section{Experimental Setup}
\subsection{Environments}
We evaluate our approach on two multi-agent environments where coordination on shared conventions is critical. 

\begin{figure}[t]
    \centering
    \includegraphics[width=0.45\columnwidth]{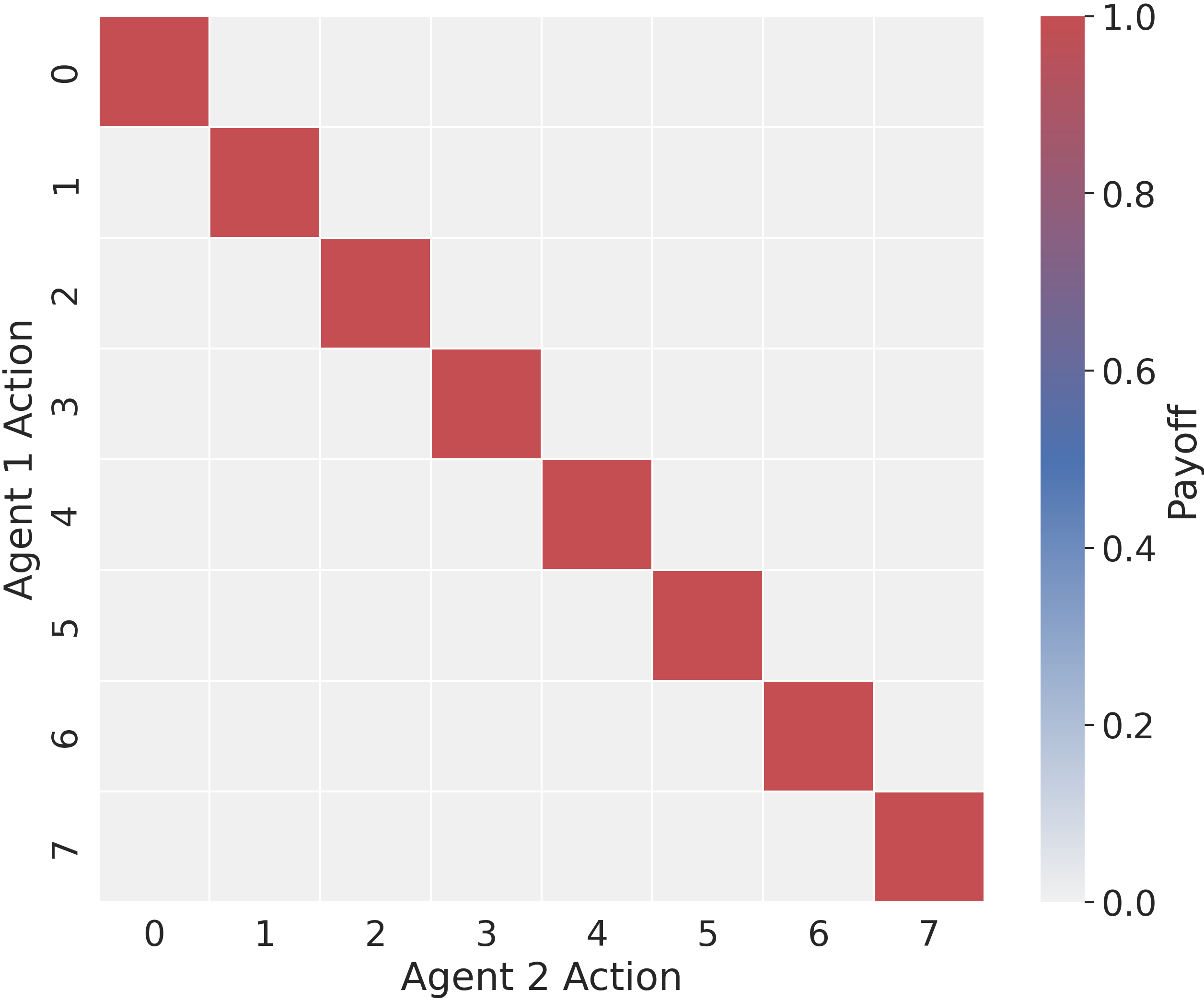}
    \hfill
    \includegraphics[width=0.45\columnwidth]{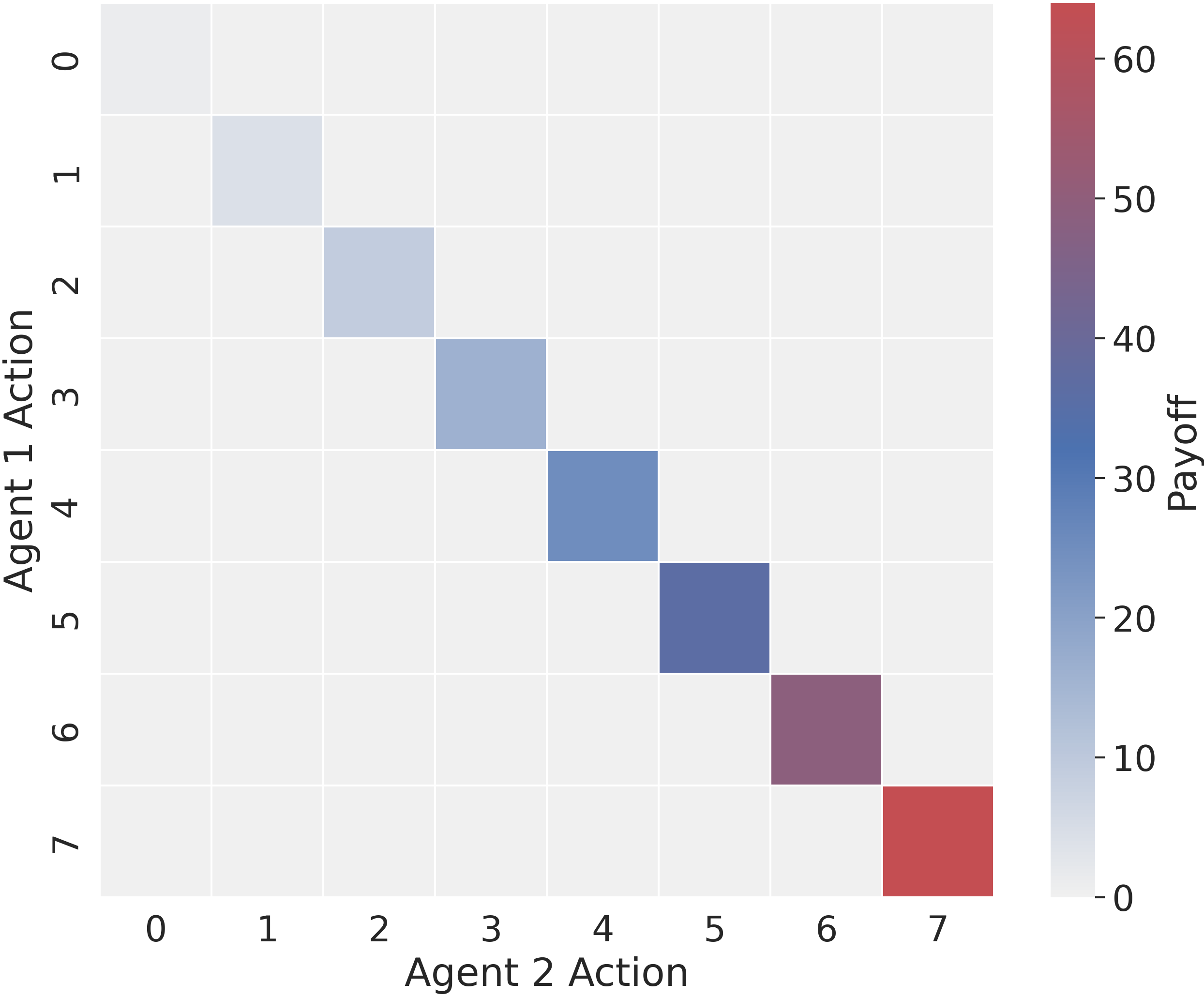}
    \caption{Payoff matrices for the Repeated Matrix Game. Left: Uniform payoffs across conventions. Right: Differentiated payoffs across conventions.}
    \label{fig:matrix_payoffs}
\end{figure}

\begin{figure}[t]
    \centering
    \includegraphics[width=0.49\columnwidth]{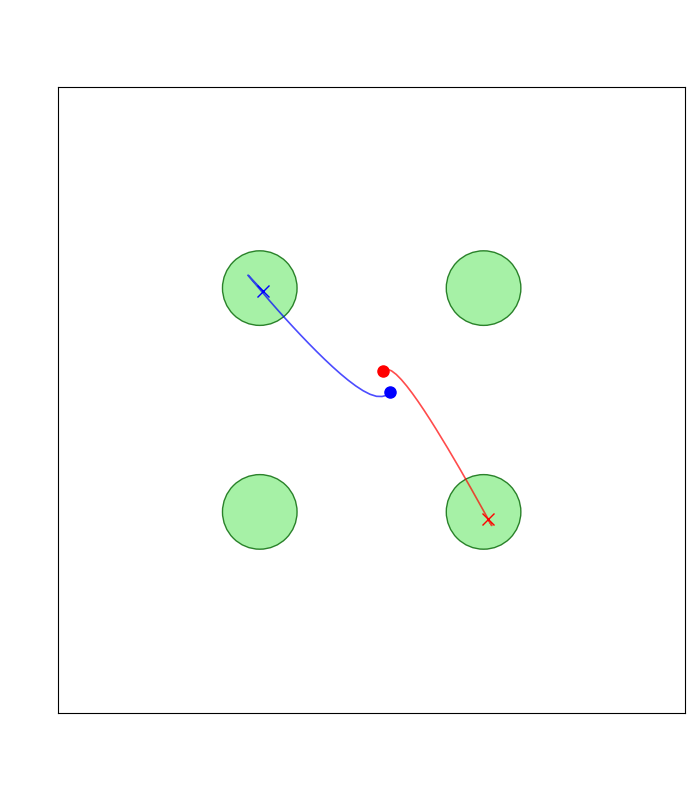}
    \hfill
    \includegraphics[width=0.49\columnwidth]{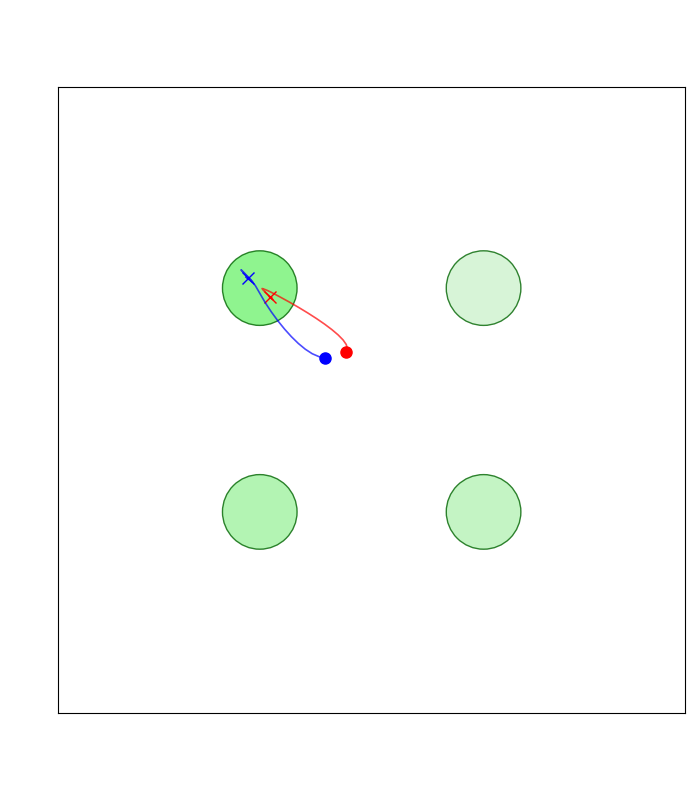}
    \caption{Trajectories of agents playing two variants of the point-mass-rendezvous game. Left: Layout with 4 landmarks with equal value; no reward is achieved as agents navigated to different landmarks. Right: 4 landmarks with differentiated rewards, visualized by opacity; both agents successfully navigating to the landmark with the highest rewards.}
    \label{fig:pmr_setup}
\end{figure}

The first is a \textbf{Repeated Matrix Game}, a canonical task where agents must select matching actions across multiple trials. We consider two distinct configurations of this game, illustrated by the payoff matrices in Figure~\ref{fig:matrix_payoffs}. In the first setting, all conventions are of equal value and are equally difficult to discover. In the second setting, conventions possess distinct values and present varying difficulties for discovery, incentivizing agents not just to find any solution but an optimal one.

The second is \textbf{Point Mass Rendezvous} (PMR) \cite{charakorn2023generating}, a time-extrapolated version of the repeated matrix game, similar to cooperative reaching \cite{rahman2023generating}. In this environment, agents must rendezvous at one of a set of pre-configured landmarks, as shown in Figure~\ref{fig:pmr_setup}. The value of each landmark may differ, with some landmarks being more valuable than others. Agents observe their position, velocity, distance to each of the landmarks, and their partner's velocity. As with the matrix game, we consider both uniform and differentiated reward structures for this task, testing the agent's ability to coordinate on both arbitrary and value-driven conventions.

\subsection{Evaluation Protocol}
Our evaluation protocol is designed to test ad-hoc teamwork capabilities across the cognitive hierarchy.  For each environment, we construct a benchmark set of hard-coded policies that align with the convention system observed in the learned $K_0$ population. We denote this evaluation set as $K_{0, \text{test}}$. Following our stratified sampling approach (Section~\ref{subsec:stratified_sampling}), we then generate a $K_{1, \text{test}}$ population by training best responses against various subsets of $K_{0, \text{test}}$. Success against this adaptive evaluation set requires the agent to actively probe an unseen partner and identify the specific conventions they are capable of supporting.

We benchmark \textit{ConventionPlay} against three baseline methods. First, a \textbf{Best Response (BR)} is trained against the $K_0$ population to establish a baseline for simple adaptation. Second, \textbf{FCP} \cite{strouseFCPCollaboratingHumans2021} introduces behavioral diversity through intermediate checkpoints; we implement this by wrapping the final $K_0$ agents in an $\epsilon$-greedy policy with varying exploration rates. Finally, \textbf{SyKLRBR} \cite{cuiSyKLRBRKlevelReasoning2021} represents state-of-the-art cognitive hierarchy modeling for zero-shot coordination; for structural parity, we use a two-level SyKLRBR hierarchy grounded in our learned $K_0$ population. All adaptive agents are trained using PPO with a recurrent architecture to manage the inherent partial observability of the coordination task. 

%%%%%%%%%%%%%%%%%%%%%%%%%%%%%%%%%%%%%%%%%%%%%%%%%%%%%%%%%%%%%%%%%%%%%%%%
\section{Results}

\begin{table*}[t]
\centering
\caption{Coordination Efficiency $\eta$ of ZSC methods across four domains. Results show mean efficiency $\pm$ standard deviation across three random seeds. \textit{ConventionPlay} demonstrates parity with partners that play with fixed conventions ($K_0$) and superior performance when paired with adaptive ($K_1$) partners.}
\label{tab:aggregated_results}

% --- Row 1 ---
\begin{minipage}{0.48\textwidth}
\centering
\textbf{Matrix Game (Uniform)}\\[0.5em]
\resizebox{\linewidth}{!}{%
\begin{tabular}{lccc}
\toprule
Method & $K_0$ Test & $K_1$ Test & Self-Play \\ \midrule
BestResponse & $\bm{77.50 \pm 0.00}$ & $85.83 \pm 3.19$ & $78.75 \pm 8.69$ \\
FCP & $61.25 \pm 5.97$ & $72.81 \pm 2.21$ & $82.36 \pm 3.57$ \\
SyKLRBR & $77.50 \pm 0.00$ & $87.19 \pm 2.90$ & $\bm{90.00 \pm 2.38}$ \\
ConventionPlay & $77.29 \pm 0.15$ & $\bm{89.79 \pm 1.41}$ & $83.06 \pm 3.36$ \\
\bottomrule
\end{tabular}

}
\end{minipage}
\hfill
\begin{minipage}{0.48\textwidth}
\centering
\textbf{Matrix Game (Differentiated)}\\[0.5em]
\resizebox{\linewidth}{!}{%
\begin{tabular}{lccc}
\toprule
Method & $K_0$ Test & $K_1$ Test & Self-Play \\ \midrule
BestResponse & $47.81 \pm 0.00$ & $65.41 \pm 0.26$ & $70.55 \pm 2.01$ \\
FCP & $47.29 \pm 3.58$ & $64.40 \pm 3.75$ & $79.29 \pm 0.95$ \\
SyKLRBR & $41.67 \pm 0.15$ & $59.06 \pm 0.29$ & $\bm{91.65 \pm 2.38}$ \\
ConventionPlay & $\bm{59.69 \pm 2.27}$ & $\bm{75.23 \pm 0.51}$ & $60.36 \pm 1.28$ \\
\bottomrule
\end{tabular}

}
\end{minipage}

\vspace{1.5em}

% --- Row 2 ---
\begin{minipage}{0.48\textwidth}
\centering
\textbf{PMR (Uniform)}\\[0.5em]
\resizebox{\linewidth}{!}{%
\begin{tabular}{lccc}
\toprule
Method & $K_0$ Test & $K_1$ Test & Self-Play \\ \midrule
BestResponse & $98.68 \pm 0.08$ & $\bm{99.00 \pm 0.05}$ & $99.84 \pm 0.07$ \\
FCP & $96.27 \pm 0.88$ & $98.08 \pm 0.15$ & $97.59 \pm 0.63$ \\
SyKLRBR & $\bm{98.89 \pm 0.19}$ & $98.93 \pm 0.17$ & $99.80 \pm 0.16$ \\
ConventionPlay & $98.63 \pm 0.10$ & $98.75 \pm 0.11$ & $\bm{99.96 \pm 0.11}$ \\
\bottomrule
\end{tabular}
}
\end{minipage}
\hfill
\begin{minipage}{0.48\textwidth}
\centering
\textbf{PMR (Differentiated)}\\[0.5em]
\resizebox{\linewidth}{!}{%
\begin{tabular}{lccc}
\toprule
Method & $K_0$ Test & $K_1$ Test & Self-Play \\ \midrule
BestResponse & $92.55 \pm 6.20$ & $78.52 \pm 3.14$ & $59.09 \pm 3.47$ \\
FCP & $94.43 \pm 2.78$ & $74.23 \pm 2.92$ & $49.45 \pm 3.63$ \\
SyKLRBR & $94.32 \pm 1.93$ & $81.06 \pm 1.46$ & $63.85 \pm 4.34$ \\
ConventionPlay & $\bm{96.05 \pm 0.44}$ & $\bm{88.82 \pm 2.22}$ & $\bm{65.43 \pm 2.36}$ \\
\bottomrule
\end{tabular}
}
\end{minipage}

\end{table*}

\begin{figure*}[t]
    \centering
    \includegraphics[width=0.9\textwidth]{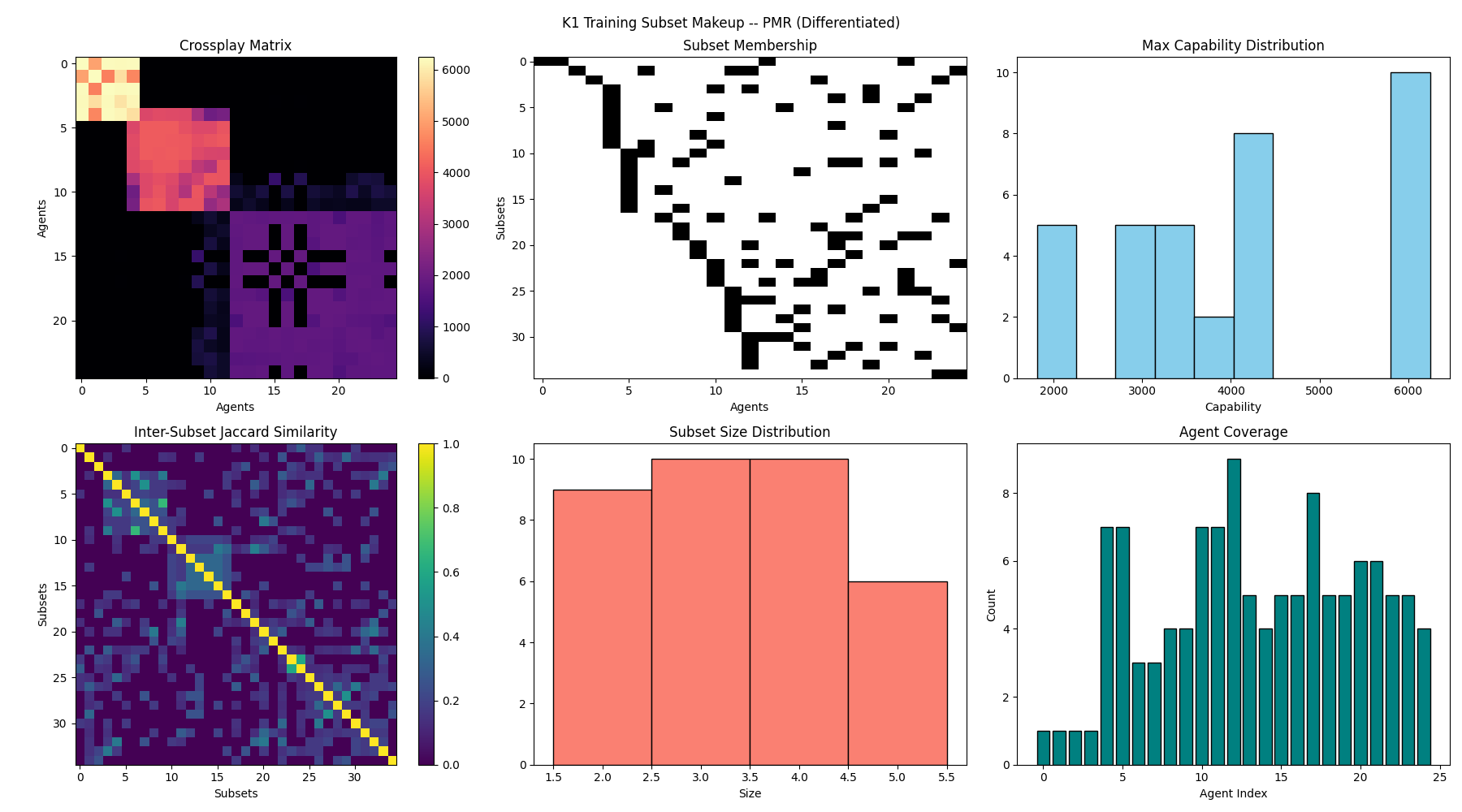}
    \caption{Makeup of the generated $K_1$ subsets. The top left image has cross-play of all the $K_0$ population; clusters here roughly visualize compatible policies which can be considered conventions. The top middle shows how the stratified sampling has placed different agents in different clusters. The top right shows the maximum capability distribution across subsets, ensuring various $K_1$ subsets have different capabilities. Bottom left shows Jaccard similarity between all generated clusters. Bottom right shows enumerated subset sizes and the number of times a specific agent is used in the support of the entire $K_1$ population.}
    \label{fig:subset_makeup}
\end{figure*}

We begin by examining the generation of the $K_0$ and $K_1$ subsets, which form the training distribution for our adaptive agent; this is illustrated in Figure~\ref{fig:subset_makeup}. The cross-play matrix of the $K_0$ population (top-left) reveals distinct clusters of mutually compatible (and incompatible) policies. Our stratified sampling process (Section~\ref{subsec:stratified_sampling}) successfully constructs a diverse population of $K_1$ agents, ensuring a broad distribution of maximum capabilities across all $K_1$ training sets (top-right). The low Jaccard similarity between subsets (bottom-left) and the broad agent coverage (bottom-right) illustrates that the $K_1$ population is exposed to a wide variety of behaviors without being redundant, or dominated by a single convention.

\begin{figure}[ht]
    \centering
    \includegraphics[width=0.5\columnwidth]{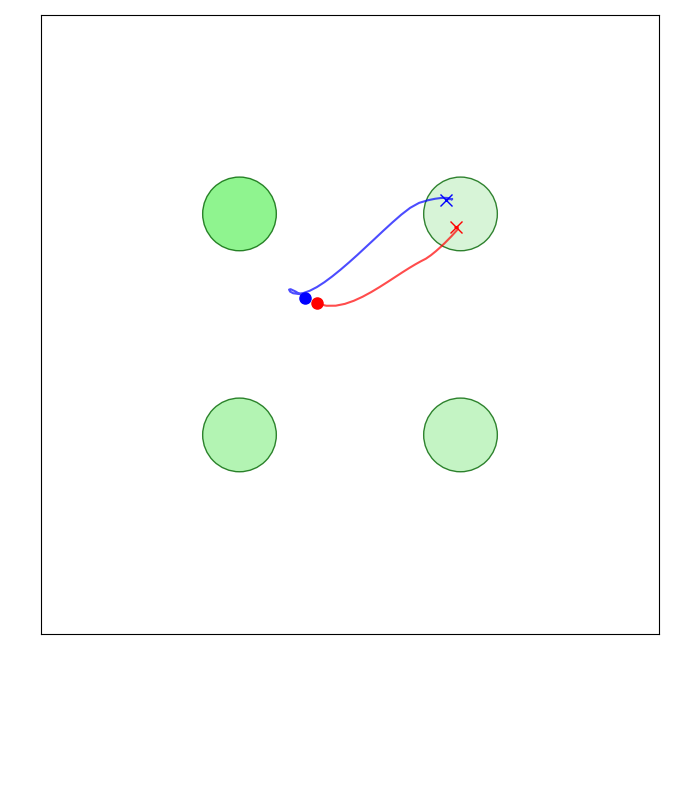}%
    \hfill
    \includegraphics[width=0.5\columnwidth]{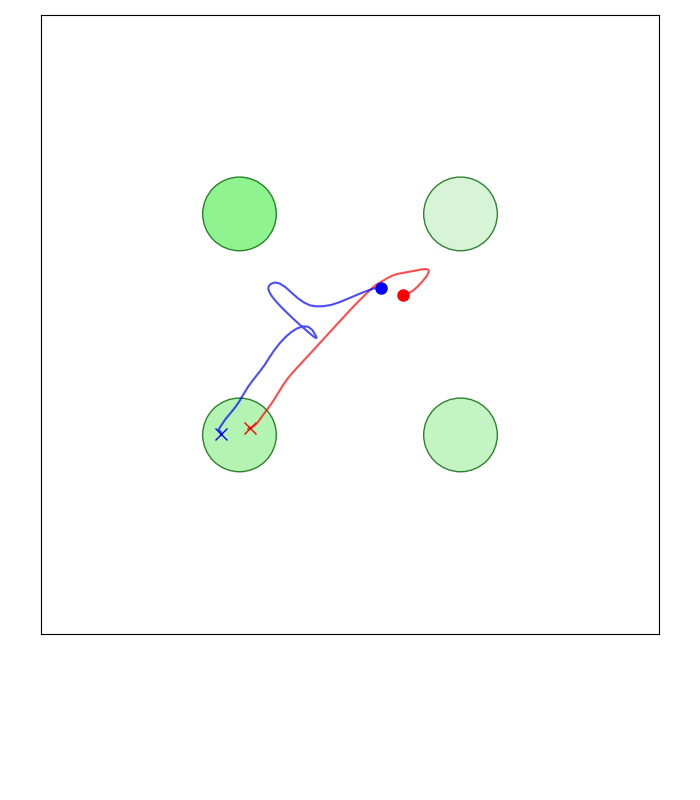}%
    \caption{Visualizing the steering behavior of \textit{ConventionPlay} (blue) against $K_0$ (left) and $K_1$ (right) partners. In the $K_0$ case, the agent probes for a better convention but converges on the partner's choice when no adaptation is detected. Conversely, with a $K_1$ partner, the agent persists in moving toward the highest value goal, then the second highest value goal, successfully influencing the adaptive partner to switch conventions.}
    \label{fig:pmr_probe}
\end{figure}

We evaluate the performance of our generalist $K_2$ agent (\textit{ConventionPlay}) against baselines. Table~\ref{tab:aggregated_results} presents the results across the evaluation domains. \textit{ConventionPlay} shows parity with baselines for uniform games and in interaction with $K_0$ partners, demonstrating that our hierarchical training does not degrade robust convention adherence. The critical difference emerges in the $K_1$ regime for differentiated games. In these settings, baseline agents lack the probing mechanisms required to identify the extent of a partner's flexibility. When paired with an adaptive $K_1$ follower, baselines often settle for the first discovered convention, even if a higher-value strategy exists within the partner's capability set $\mathcal{K}_{\pi_1}$.
In contrast, \textit{ConventionPlay} achieves significantly higher coordination efficiency $\eta$ by actively steering the interaction. As visualized in Figure~\ref{fig:pmr_probe}, when the agent detects that a partner is capable of adaptation, it persists in signaling high-value conventions rather than immediately conforming to the partner's initial trajectory. This "probing" behavior allows the team to converge on the optimal shared strategy allowed by the partner's repertoire. 

%%%%%%%%%%%%%%%%%%%%%%%%%%%%%%%%%%%%%%%%%%%%%%%%%%%%%%%%%%%%%%%%%%%%%%%%
\section{Discussion and Future Work}

In this work, we have introduced \textit{ConventionPlay} as a method for robust ad-hoc collaboration. By introducing strategic diversity into the $K_1$ layer of a cognitive hierarchy, we have shown that agents can learn to dynamically navigate the balance between leading and following.  The primary limitation in our method is the reliance on $J_{SP}$ for capability-aware sampling, which lacks discriminative power when distinct conventions yield similar returns. This can limit the variety of the $K_1$ population by failing to distinguish between strategically diverse behaviors. Integrating more granular metrics, such as trajectory diversity \cite{lupuTrajectoryDiversityZeroShot2021}, would allow \textit{ConventionPlay} to model different types of generalist followers.

While our current evaluation focuses on cooperative settings, the principles of \textit{ConventionPlay} could be extended to mixed-motive settings as well. In scenarios where agents have partially aligned incentives or private utilities, the ability to probe a partner's behavior and steer the interaction toward a mutually beneficial convention becomes even more critical for preventing sub-optimal outcomes.

%%%%%%%%%%%%%%%%%%%%%%%%%%%%%%%%%%%%%%%%%%%%%%%%%%%%%%%%%%%%%%%%%%%%%%%%
%%% The next two lines define, first, the bibliography style to be 
%%% applied, and, second, the bibliography file to be used.
\bibliographystyle{ACM-Reference-Format} 
\bibliography{all_items}

%%%%%%%%%%%%%%%%%%%%%%%%%%%%%%%%%%%%%%%%%%%%%%%%%%%%%%%%%%%%%%%%%%%%%%%%
\end{document}